\documentclass[aps, a4paper,superscriptaddress,12pt,preprintnumbers,floatfix,nofootinbib]{revtex4-1}
\usepackage{url}
\usepackage{xcolor}
\usepackage{hyperref}
\usepackage[normalem]{ulem}
\hypersetup{colorlinks=true,linkcolor=redLinks,citecolor=greenLinks,
urlcolor=redLinks,
pdfborder={0 0 1}}
\hypersetup{
    bookmarks=true,         
    unicode=false,          
    pdftoolbar=true,        
    pdfmenubar=true,        
    pdffitwindow=false,     
    pdfstartview={FitH},    
    pdftitle={My title},    
    pdfauthor={Author},     
    pdfsubject={Subject},   
    pdfcreator={Creator},   
    pdfproducer={Producer}, 
    pdfkeywords={keyword1, key2, key3}, 
    pdfnewwindow=true,      
    colorlinks=false,       
    linkcolor=red,          
    citecolor=green,        
    filecolor=magenta,      
    urlcolor=cyan           
}
\colorlet{shadecolor}{gray!15}
\definecolor{greenLinks}{rgb}{0, 0.6, 0} 
\definecolor{blueLinks}{rgb}{0, 0, 0.6}
\definecolor{redLinks}{rgb}{0.6, 0, 0}
\definecolor{tempText}{rgb}{0.55, 0.10,0.67}
\definecolor{eprintLinks}{rgb}{0.4, 0.4, 0.4}
\definecolor{journalLinks}{rgb}{0.6, 0, 0}
\newcommand{\MYhref}[3][redLinks]{\href{#2}{\color{#1}{#3}}}%
\usepackage{subfigure}
\usepackage{rotating}
\usepackage{slashed}     

\usepackage[T1]{fontenc} 
\usepackage[utf8]{inputenc}
\usepackage[english]{babel}
\usepackage{amsmath}
\usepackage{graphicx}
\usepackage{color}
\usepackage{mathrsfs}   
\usepackage{amssymb}
\usepackage{hyperref}
\usepackage{dcolumn}
\usepackage{bm}
\usepackage{graphicx}

\def\lnv{lepton number violation }
\def\vev#1{\left\langle #1\right\rangle}
\newcommand{\sm}{Standard Model }
\def\SM{$\mathrm{SU(3)_c \otimes SU(2)_L \otimes U(1)_Y}$ }
 
\def\TrTrOne{ $\mathrm{SU(3)_c \otimes SU(3)_L \otimes U(1)_X}$ }
\newcommand{\AddrAHEP}{
  {\it AHEP Group, Instituto de F\'{\i}sica Corpuscular --
    C.S.I.C./Universitat de Val{\`e}ncia \\
    Edificio de Institutos de Paterna,
 C/Catedratico Jos\'e Beltr\'an, 2 \\E-46980 Paterna (Val\`{e}ncia) - SPAIN}}

\begin{document}
\title{Dirac neutrinos and dark matter stability from lepton
  quarticity} \author{Salvador Centelles Chuli\'a}
\email{salcen@alumni.uv.es} \affiliation{\AddrAHEP} \author{Ernest Ma}
\email{ma@phyun8.ucr.edu} \affiliation{Physics Department, University
  of California, Riverside, California 92521, USA} \author{Rahul
  Srivastava} \email{rahuls@prl.res.in} \affiliation{Physical Research
  Laboratory, Navrangpura, Ahmedabad - 380 009, INDIA} \author{Jos\'e
  W.F. Valle} \email{valle@ific.uv.es} \affiliation{\AddrAHEP}
\vspace{1cm} 
\begin{abstract}
   \vspace{1cm}
   
   We propose to relate dark matter stability to the possible Dirac
   nature of neutrinos. The idea is illustrated in a simple scheme where
   small Dirac neutrino masses arise from a type--I seesaw mechanism
   as a result of a $Z_4$ discrete lepton number symmetry.
   The latter implies the existence of a viable WIMP dark matter
   candidate, whose stability arises from the same symmetry which
   ensures the Diracness of neutrinos.
   
   \end{abstract}

\maketitle


 \section{Introduction}
\label{sec:introduction}


Amongst the major shortcomings of the \sm are the neutrino mass and
the dark matter problems. Underpinning the origin of neutrino mass and
elucidating the nature of dark matter would constitute a gigantic step
forward in particle physics.
Here we focus on the possibility that the neutrino mass and dark
matter problems may be closely 
interconnected~\cite{Valle:2012yx,Lattanzi:2014mia}.
Concerning neutrino mass a major unknown is whether neutrinos are
their own anti--particles, an issue which has remained an open
challenge ever since Ettore Majorana's pioneer idea on the quantum
mechanics of spin.
On the other hand, since many years physicists have pondered about
what is the dark matter made of, and what makes it stable, a property
usually assumed in an \textit{ad-hoc} fashion~\footnote{Attempts at
  stabilizing dark matter by using the $Z_2$ and $Z_3$ symmetries have
  already been considered in the
  literature~\cite{PhysRevLett.115.011801,Ma:2015mjd}.  }.
Indeed, although the existence of non-baryonic dark matter is well
established by using cosmological and astrophysical probes, its nature
has otherwise remained elusive~\cite{Bertone:2010}.

The detection of neutrinoless double beta decay would be a major step
in particle physics since, according to the black--box
theorem~\cite{Schechter:1981bd,Duerr:2011zd} it would not only
demonstrate that lepton number is violated in nature, but also imply
that neutrinos are of Majorana type.
On the other hand, the fact that the weak interaction is V-A turns
this quest into a major
challenge~\cite{avignone:2007fu,Barabash:2004pu}.
As of now the nature of neutrinos remains as mysterious as the
mechanism responsible for generating their small masses. Little is
known regarding the nature of its associated messenger particles, the
underlying mass scale or its flavor structure~\cite{Valle:2015pba},
currently probed only in neutrino oscillation
experiments~\cite{Maltoni:2004ei}.
Here we assume that neutrinos are Dirac particles. Realizing this
possibility requires extra assumptions beyond the standard \SM
electroweak gauge invariance.
One approach is to extend the gauge group itself, for example, by
using the \TrTrOne gauge structure due to its special
features~\cite{Singer:1980sw}. In this framework it has recently been
shown how to obtain a type-II seesaw mechanism for Dirac
neutrinos~\cite{Valle:2016kyz, Addazi:2016xuh}.
Using unconventional lepton charges for right handed
neutrinos~\cite{peltoniemi:1993ss} and gauging $B-L$ can also lead to
Dirac neutrinos within the type-I seesaw mechanism \cite{Ma:2014qra,
  Ma:2015raa, Ma:2015mjd}.
Alternatively, one may stick to the \SM gauge structure but use extra
flavor symmetries implying a conserved lepton number, so as to obtain
Dirac neutrinos, as suggested in~\cite{Aranda:2013gga}.

In this letter we focus on having neutrinos as Dirac particles as a
result of the cyclic symmetry $Z_4$, which plays the role of a
discrete version of lepton number, we call \textit{quarticity}. We
show that a WIMP dark matter candidate can naturally emerge,
stabilized by quarticity, the same symmetry associated to the
Diracness of neutrinos.
In Sect.~\ref{sec:model-} we sketch in some detail the required
particle content of the model and demonstrate how the Dirac nature of
neutrinos and the stability of dark matter follow from the same
principle.
In Sect.~\ref{sec:dark-matter} we discuss the main aspects concerning
our WIMP dark matter candidate, including a brief discussion of the
interactions relevant for determining its relic density. We also
discuss its direct detection potential through the Higgs portal
mechanism.  Finally we summarize our results in
Sect.~\ref{sec:summary-conclusions-}.


 \section{The Model }
\label{sec:model-}


Our model is based on the discrete symmetry $Z_4 \otimes Z_2$ where
$Z_4$ is the cyclic group of order four and $Z_2$ is the cyclic group
of order two. The group $Z_4$ can be viewed as a discrete version of
lepton number. As we show here, $Z_4$ symmetry not only forbids
Majorana terms but also forbids couplings of potential dark matter
candidate ensuring stability. Thus, the same symmetry which implies
the Dirac nature of neutrinos also ensures the stability of dark
matter.  The $Z_2$ symmetry is required here to ensure the seesaw
origin of neutrino mass, by forbidding tree level coupling between the
left and right handed neutrinos.\\

The particle content of our model along with the $Z_4$ charge
assignments of the particles are as shown in Table \ref{tab1}.
\begin{table}[h]
\begin{center}
\begin{tabular}{c c c || c c c}
  \hline \hline
  Fields           \hspace{1cm}        &  $Z_4$               \hspace{1cm}          & $Z_2$                \hspace{1cm}         & 
  Fields           \hspace{1cm}        &  $Z_4$               \hspace{1cm}          & $Z_2$                                     \\
  \hline \hline
  $\bar{L}_{i,L}$  \hspace{1cm}        &  $\mathbf{z}^3$      \hspace{1cm}          & $\mathbf{1}$          \hspace{1cm}        &  
  $\nu_{i,R}$      \hspace{1cm}        &  $\mathbf{z}$        \hspace{1cm}          & $\mathbf{-1}$                              \\
  $ l_{i,R} $      \hspace{1cm}        &  $\mathbf{z}$        \hspace{1cm}          & $\mathbf{1}$          \hspace{1cm}        & 
  $\bar{N}_{i,L}$  \hspace{1cm}        &  $\mathbf{z}^3$      \hspace{1cm}          & $\mathbf{1}$                                \\
  $N_{i,R}$        \hspace{1cm}        &  $\mathbf{z}$        \hspace{1cm}          & $\mathbf{1}$         \hspace{1cm}         &                                      
                   \hspace{1cm}        &                      \hspace{1cm}          &                                             \\
  \hline \hline
  $\Phi$         \hspace{1cm}        &  $\mathbf{1}$        \hspace{1cm}          & $\mathbf{1}$         \hspace{1cm}         &
  $\chi$         \hspace{1cm}        &  $\mathbf{1}$        \hspace{1cm}          & $\mathbf{-1}$                               \\
  $\zeta$          \hspace{1cm}        &  $\mathbf{z}$        \hspace{1cm}          & $\mathbf{1}$          \hspace{1cm}        &          
  $\eta$           \hspace{1cm}        &  $\mathbf{z}^2$      \hspace{1cm}          & $\mathbf{1}$                               \\
    \hline \hline
  \end{tabular}
\end{center}
\caption{ 
  The $Z_4$ and $Z_2$ charge assignments for leptons and the scalars  
  ($\Phi$, $\chi$, $\zeta$ and $\eta$). 
  Here $\mathbf{z}$ denotes the fourth root of unity, i.e. $\mathbf{z}^4 = 1$.}
  \label{tab1}
\end{table}

In Table \ref{tab1}, $L_{i,L} = (\nu_{i,L}, l_{i,L})^T$; $i = e, \mu,
\tau$ are the lepton doublets which have charge $\mathbf{z}$ under
$Z_4$.  The $l_{i,R}$; $i = e, \mu, \tau$ are the charged lepton
singlets which carry $Z_4$--charge $\mathbf{z}$.
Apart from the \sm fermions, the model also includes three
right--handed neutrinos $\nu_{i,R}$ transforming as singlets under the
\SM gauge group and with charge $\mathbf{z}$ under $Z_4$. We also add
three gauge singlet Dirac fermions $N_{i,L}, N_{i,R}$; $i = 1, 2, 3$
with charge $\mathbf{z}$ under $Z_4$, as shown in
Table.~\ref{tab1}. All the fermions except $\nu_{i,R}$ are even under
the $Z_2$ symmetry, while $\nu_{i,R}$ are odd.
 
In the scalar sector the $\Phi = (\phi^+, \phi^0)^T$ transforms as
$\mathrm{SU(2)_L}$ doublet, does not carry any $Z_4$ charge and is
even under $Z_2$ symmetry.
On the other hand the real scalar $\chi$ is a gauge singlet, carries
no $Z_4 $ charge but is odd under $Z_2$ symmetry.  We also add two
other gauge singlet scalars $\zeta$ and $\eta$ both of which carry
$Z_4$ charges $\mathbf{z}$ and $\mathbf{z}^2$ respectively and are
even under $Z_2$. The fact that $\mathbf{z}^2 = -1$ allows us to take
the field $\eta$ also to be real.\\
 
\begin{figure}[!h]
\centering
\includegraphics[width=.6\textwidth]{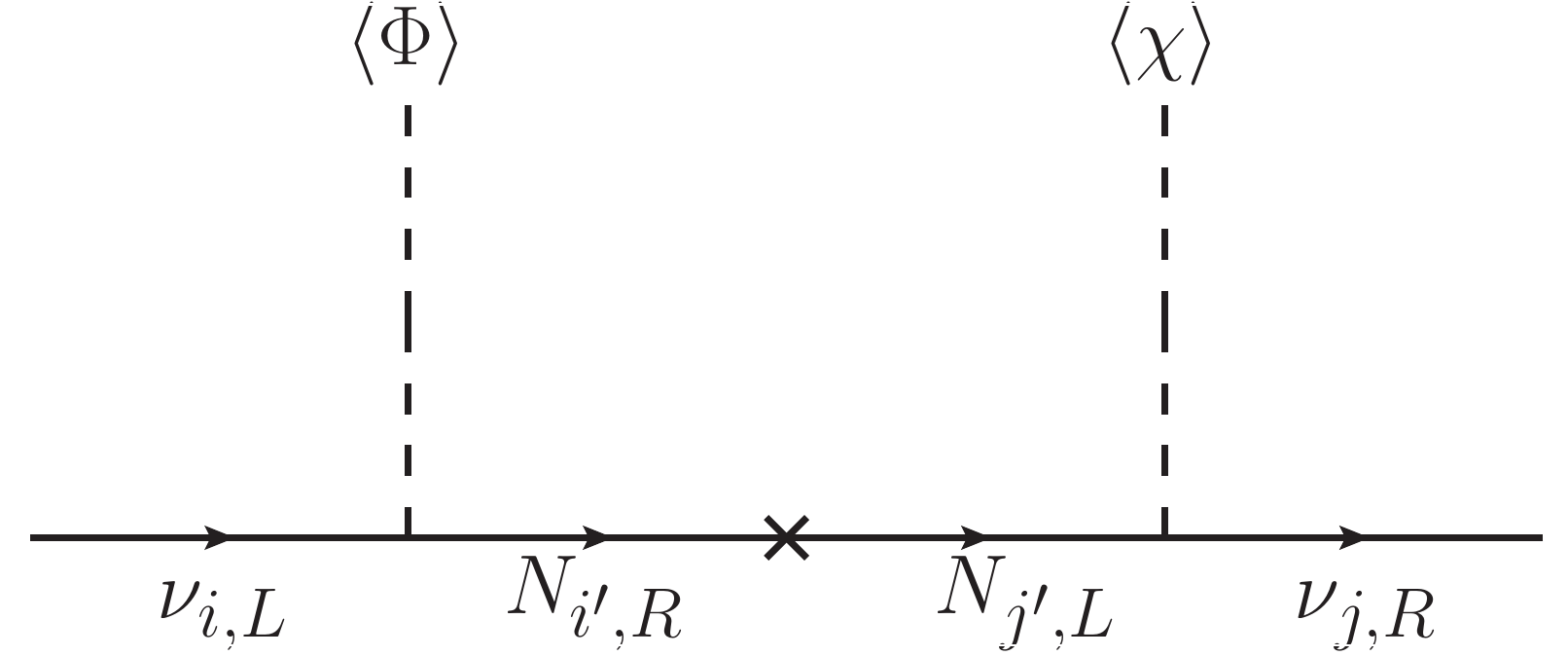}
\caption{Dirac type-I seesaw mechanism.  }
\label{feyn}
\end{figure}

The \SM $ \otimes$ $Z_4 \otimes Z_2$ invariant Yukawa Lagrangian for
the leptons is given by
 \begin{eqnarray}
 - \mathcal{L}_{\rm{Yuk}, l}  & = & y^l_{ij} \, \bar{L}_{i,L} \, \Phi \, l_{j, R}
  \, + \, f_{ij} \, \bar{L}_{i,L} \, \tilde{\Phi} \, N_{j, R} \, + \,  g_{ij} \, \bar{N}_{i,L} \, \chi \, \nu_{j, R} 
 \, + \, M_{ij} \,\bar{N}_{i,L} \,N_{j, R}   \nonumber \\
 & + & y^\eta_{ij} \, \bar{\nu}^c_{i,R} \, \nu_{j,R} \, \eta \, \,  + \, \, \rm{h.c.}
   \label{yuk}
 \end{eqnarray}
 where $y_{ij}, f_{ij}, g_{ij}, y^\eta_{ij}$, $i,j = 1, 2, 3$, are the
 Yukawa couplings. Also $\tilde{\Phi}$ is the
 $\mathrm{SU(2)_L}$ conjugate field, $\tilde{\Phi} = i \sigma_2
 \Phi^*$ where $\sigma_2$ is the second Pauli matrix. After symmetry
 breaking the $Z_4$ neutral scalars $\Phi, \chi$ acquire vacuum
 expectation values (vevs) $$\vev{ \Phi} = \frac{v}{\sqrt{2}}~~~{\rm and}~~~ \vev{\chi}
 = u.$$ The $Z_4$ charged scalars $\eta, \zeta$ acquire no vev thus
 ensuring that the $Z_4$ symmetry remains unbroken even after
 electroweak symmetry breaking. Since the scalar $\chi$ was odd under
 $Z_2$ symmetry, its vev breaks the $Z_2$ symmetry spontaneously. The
 neutrinos can then acquire a tiny mass through type-I Dirac seesaw
 mechanism as shown in Fig. \ref{feyn} and briefly discussed below.
 
 After symmetry breaking the charged leptons acquire mass through the
 vev of the $\mathrm{SU(2)_L}$ doublet scalar $\Phi$. The $6 \times 6$
 mass matrix for the neutrinos and heavy fermions $N_{i,L}, N_{i,R}$
 in the basis $(\bar{\nu}_{i,L}, \bar{N}_{i,L})$ and $(\nu_{i,R},
 N_{i,R})^T$ is given by
  \begin{eqnarray}
  M_{\nu, N} & = &  \left( 
\begin{array}{cc}
0                 & f \, v/\sqrt{2}      \\
g \, u       & M                             \\
\end{array}
\right)~.
\label{seesaw}
 \end{eqnarray}
 where $f, g, M$ are each $3 \times 3$ matrices with elements $f_{ij},
 g_{ij}, M_{ij}$ respectively. The elements of the invariant mass
 matrix $M$ for the heavy leptons $N_{i,L}, N_{i,R}$ can be naturally
 much larger than the symmetry breaking scales appearing in the
 off-diagonal blocks, i.e.  $ M_{ij} \gg v, u$.
 In this limit the mass matrix in (\ref{seesaw}) can be easily block
 diagonalized by the perturbative seesaw diagonalization method given
 in~\cite{Schechter:1981cv}. The resulting $3 \times 3$ mass matrix
 for light neutrinos can be viewed as the Dirac version of the well
 known type-I seesaw mechanism and is given as follows
    \begin{eqnarray}
   M_{\nu} & = & \frac{u \, v}{\sqrt{2}} \, f \, M^{-1} \, g
   \label{numass}
  \end{eqnarray}
Note that $\Phi$ and $\chi$ are both singlets under $Z_4$. While
$\Phi$ is even under $Z_2$, $\chi$ is odd and its vev breaks $Z_2$
symmetry inducing neutrino mass.\\

Turning now to the \SM $ \otimes$ $Z_4 \, \otimes Z_2$ invariant
scalar potential, we can write it as 
\begin{eqnarray}
 V & = & - \mu^2_{\Phi} \, \Phi^\dagger \Phi  \, - \, \frac{\mu^2_\chi}{2} \, \chi^2 \, + \, \frac{m'^2_\eta}{2} \, \eta^2 \, + \, m'^2_\zeta \, \zeta^* \zeta \, + \, \frac{\kappa}{2} \, \eta \, \zeta^2 \, + \, \frac{\lambda_\Phi}{4} \, (\Phi^\dagger \Phi)^2 
 \, + \, \frac{\lambda_\chi}{16} \, \chi^4   \, + \, \frac{\lambda_\eta}{16} \, \eta^4
 \nonumber \\
& + & \frac{\lambda_\zeta}{4} \, (\zeta^* \zeta)^2 \, + \, \frac{\lambda'_\zeta}{16} \, \zeta^4  
\, + \, \frac{\lambda_{\Phi \chi}}{2} \, \Phi^\dagger \Phi \, \chi^2  \, + \, \frac{\lambda_{\Phi \eta}}{2} \, \Phi^\dagger \Phi \, \eta^2 \, + \, \lambda_{\Phi \zeta} \, \Phi^\dagger \Phi \, \zeta^* \zeta \, + \, \frac{\lambda_{\chi \eta}}{4} \, \chi^2 \, \eta^2   \nonumber \\
& + & \frac{\lambda_{\chi \zeta}}{2} \, \chi^2 \, \zeta^* \, \zeta  \, + \,
\frac{\lambda_{\zeta \eta}}{2} \, \eta^2 \, \zeta^* \, \zeta \quad + \quad \rm{h.c.} 
 \label{pot}
\end{eqnarray}

 After symmetry breaking one has, in the unitary gauge  
\begin{eqnarray}
 \Phi & \rightarrow & \frac{1}{\sqrt{2}} \left( 
\begin{array}{c}
0                 \\
 v + h'                  \\
\end{array}
\right)~,
\qquad \chi \, \rightarrow \, u + \chi'
\end{eqnarray}
The minimization conditions are given by
\begin{eqnarray}
 \mu^2_\Phi & = & \frac{\lambda_\Phi}{4} \, v^2 \, + \, \frac{\lambda_{\Phi \chi}}{2} \, u^2 \nonumber \\
 \mu^2_\chi & = & \frac{\lambda_\chi}{4} \, u^2 \, + \, \frac{\lambda_{\Phi \chi}}{2} \, v^2
\end{eqnarray}
The $h'$ and $\chi'$ fields mix with each other and their mass squared
matrix in the $(h', \chi')$ basis is given by
\begin{eqnarray}
  M^2_{h', \chi'} & = & \frac{1}{4}  \left( 
\begin{array}{cc}
 \lambda_{\Phi} v^2    &  2 \lambda_{\Phi \chi} u \, v     \\                                 
2 \lambda_{\Phi \chi} u \, v      &  \lambda_{\chi} u^2           \\
\end{array}
\right)~.
\label{smass}
\end{eqnarray}
After diagonalization the CP--even neutral mass eigenstates $h$ and
$\chi$ of \ref{smass} are given by
\begin{eqnarray}
h & = & \cos \theta \, h' \, + \, \sin \theta \, \chi' \nonumber  \\
\chi & = & -\sin \theta \, h' \, + \, \cos \theta \, \chi'
 \label{masseig}
\end{eqnarray}
where the doublet--singlet mixing angle $\tan 2 \theta =
\dfrac{4\lambda_{\Phi \chi} u \, v}{\lambda_{\Phi} v^2 -
  \lambda_{\chi} u^2}$. We identify $h$ as the 125 GeV particle
recently discovered at LHC. 

\begin{figure}[!h]
\centering
\includegraphics[width=.5\textwidth]{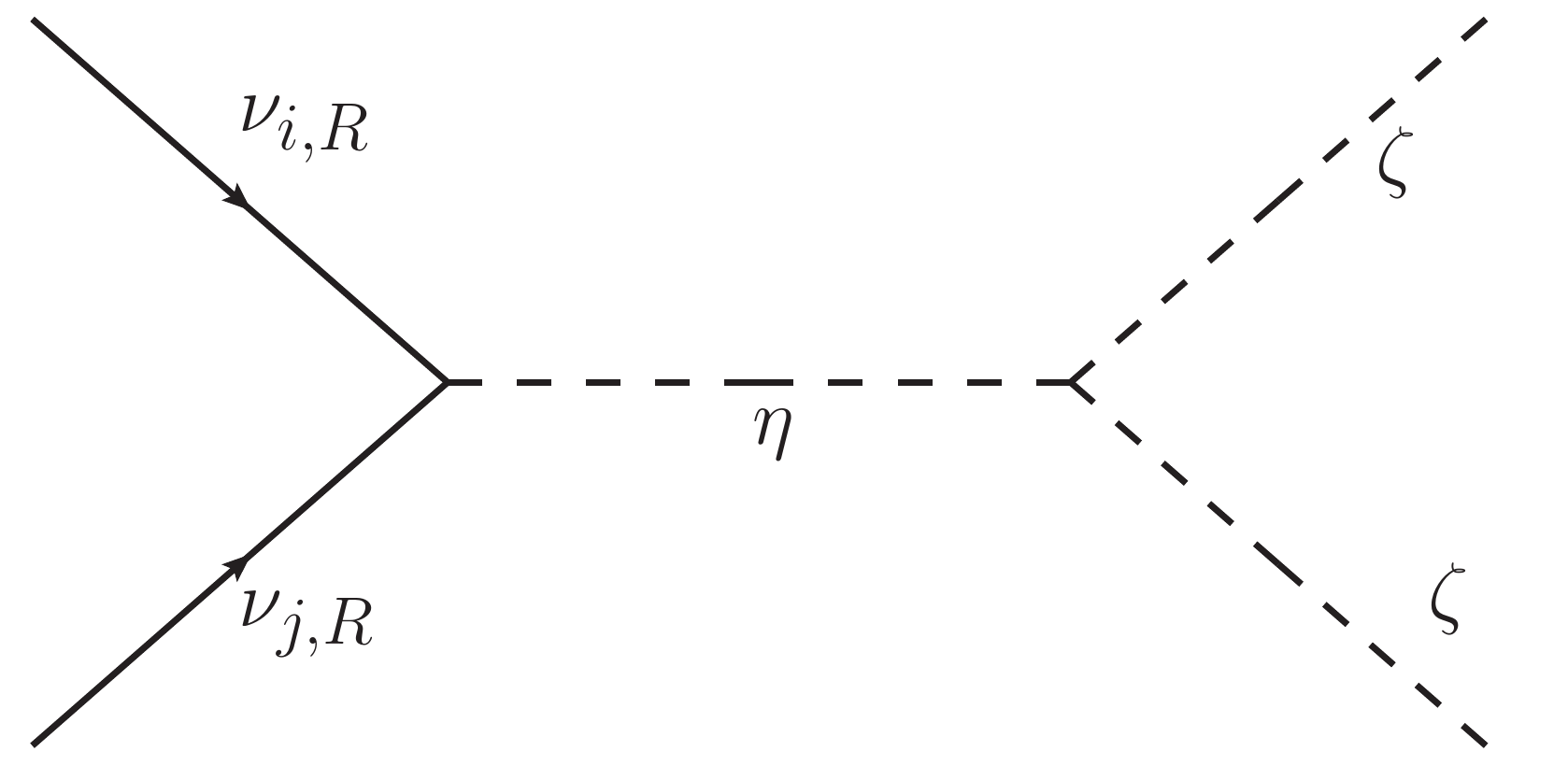}
\caption{The coupling between right handed neutrinos and $\zeta$
  mediated by the scalar $\eta$. }
\label{fig4}
\end{figure}

Let us discuss briefly the role of the $\zeta$ and $\eta$ scalars
which are singlets under the \SM gauge group, are even under $Z_2$
symmetry, but carry $Z_4$ charges $\mathbf{z}$ and $\mathbf{z}^2$
respectively.
Since $\eta, \zeta$ fields do no acquire vevs, they do not mix with
other fields. However, their masses do receive contribution from the
vev of $\Phi, \chi$ and are given by
\begin{eqnarray}
 m^2_\eta & = & m'^2_\eta \, + \, \frac{\lambda_{\Phi \eta}}{2} \, v^2  \, + \, \frac{\lambda_{\chi \eta}}{2} \, u^2 \nonumber \\
m^2_{\zeta} & = & m'^2_\zeta \, + \, \frac{\lambda_{\Phi \zeta}}{2} \, v^2  \, + \, \frac{\lambda_{\chi \zeta}}{2} \, u^2
 \label{mdm}
\end{eqnarray}
 In the absence of $\zeta$ and $\eta$, the Lagrangian of the model
 acquires an enhanced symmetry namely \SM $ \, \otimes \, U(1) \,
 \otimes \, Z_2 $ where $U(1)$ is a continuous global symmetry which
 can be interpreted as a generalized global lepton number symmetry.
In the presence of the scalars $\zeta$ and $\eta$ one can write
following $Z_4 \otimes Z_2$ invariant terms in the scalar potential
\begin{eqnarray}
 \eta^2,\, \, \eta \, \zeta^2, \, \, \eta^4, \, \, \zeta^4, \, \,  \eta^2 \, \zeta^* \, \zeta \quad + \quad \rm{H.c.} 
 \label{z4invpot}
\end{eqnarray}
Notice that all these terms are $Z_4 \otimes Z_2$ invariant but break
the global $U(1)$ invariance so the remaining symmetry group is just
$Z_4 \otimes Z_2$. Note that apart from the above terms, the other
terms in scalar potential \ref{pot} are invariant under the global
U(1).
On the other hand the field $\eta$ also couples to the right handed
neutrinos through a $Z_4 \otimes Z_2$ invariant term
\begin{eqnarray}
  {\nu}^T_{i,R} \,\sigma_2 \nu_{j,R} \, \eta \, \, + \, \, \rm{h.c.}
 \label{z4invyuk}
\end{eqnarray}
Again this Yukawa coupling is only $Z_4 \otimes Z_2$ invariant, breaking the
continuous $U(1)$ symmetry. Owing to the couplings of $\eta$ to the
scalar $\zeta$ in \ref{z4invpot} and to right handed neutrinos as in
\ref{z4invyuk}, the latter can also couple to $\zeta$ as shown in
Fig. \ref{fig4}.

After the scalars $\Phi$ and $\chi$ acquire vevs, the $Z_2$ symmetry
breaks spontaneously, implying a small mass for neutrinos. However,
since neither $\Phi$ nor $\chi$ carries a $Z_4$ charge, one has
that $Z_4$ remains exact even after spontaneous electroweak breaking.
The unbroken $Z_4$ symmetry ensures that the neutrinos do not acquire
any Majorana mass term, retaining their Dirac nature even after
spontaneous symmetry breaking takes place.

On the other hand, since the $\zeta$ and $\eta$ fields carry the $Z_4$
charge, this implies that $\zeta$ as well as $\eta$ should not acquire
any vev in order to prevent breaking $Z_4$.  This ensures that
neutrinos remain Dirac and also leads to the possibility that $\zeta$
can be a stable particle and thus a potential candidate for dark
matter. We discuss this possibility in the next section.


\section{WIMP scalar dark matter candidate}
\label{sec:dark-matter}

 
We now turn briefly to the issue of stability of $\zeta$ and its
suitability as a dark matter candidate.
As mentioned before, to ensure that neutrinos remain Dirac particles,
the $Z_4$ quarticity symmetry should remain exact, so that no scalar
carrying a $Z_4$ charge should acquire any vev.
Thus both $\eta$ and $\zeta$ which carry $Z_4$ charges can potentially
be stable as a result of the unbroken $Z_4$ symmetry. However, owing
to the $Z_4$ charge assignments, the $\eta$ field has cubic couplings
to both scalars and fermions as noted in Eqs.~\ref{z4invpot} and
\ref{z4invyuk}. These couplings lead to the decay of $\eta$ and thus,
despite carrying a $Z_4$ charge, $\eta$ is not stable.
 
On the other hand, due to its $Z_4$ charge, all cubic couplings of
$\zeta$ are forbidden by the unbroken $Z_4$ symmetry.  This implies
that there is no term of the form $\zeta \rho_i \rho_j$ where $\rho_i,
\rho_j$ stands for other scalar species of the model which is allowed
by $Z_4$.
Likewise, in the Yukawa sector the $Z_4$ symmetry also ensures that
all terms of the form $\zeta \psi_i \psi_j$, $\psi_i$ denoting a
generic fermion, are forbidden. Thus, the residual $Z_4$ symmetry
responsible for the Dirac nature of neutrinos also ensures the
stability of the $\zeta$ making it a potentially viable dark matter
candidate.
\begin{figure}[!h]
\centering
\includegraphics[width=.4\textwidth]{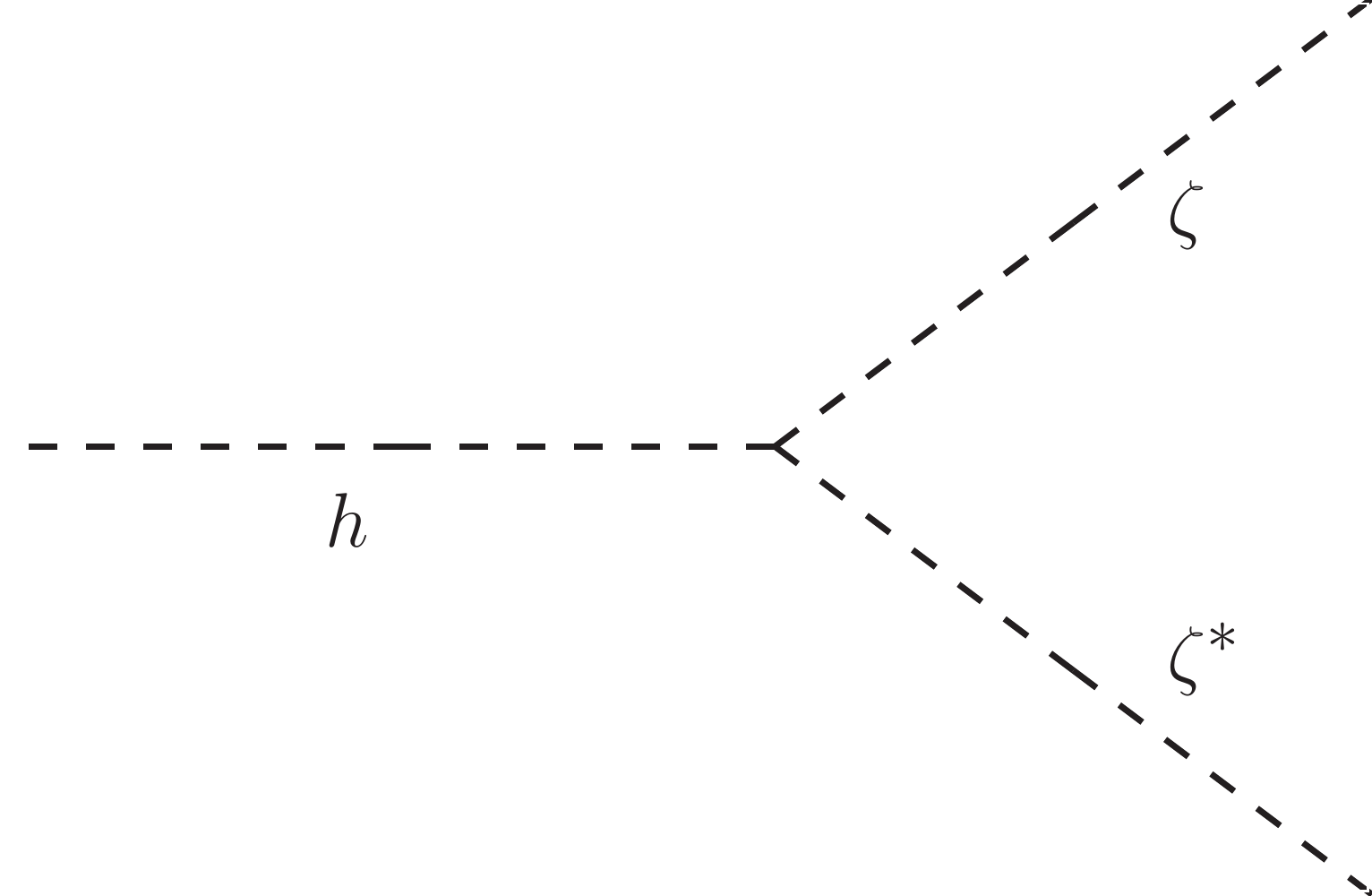} \hspace{2cm}
\includegraphics[width=.4\textwidth]{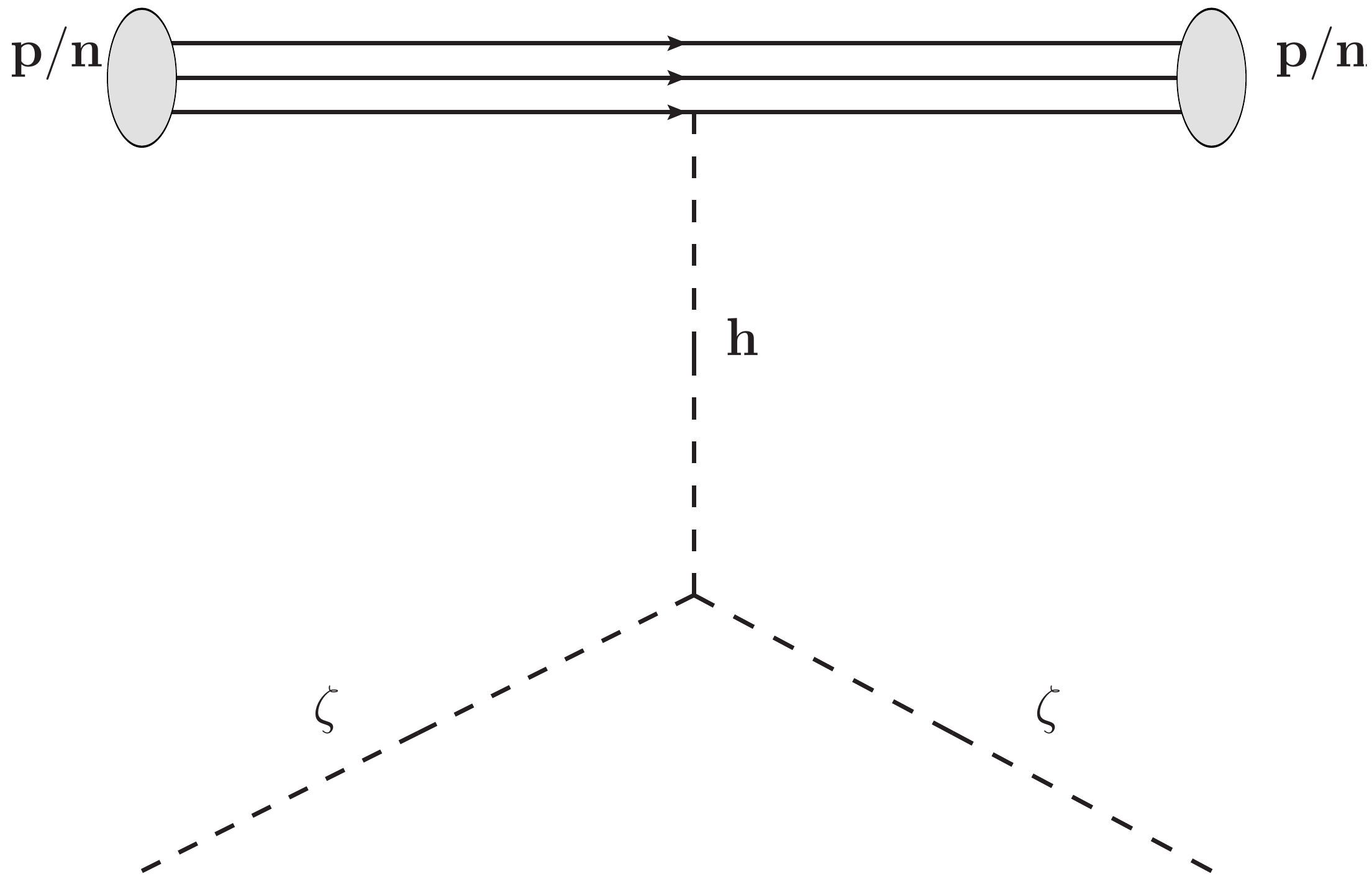}
\caption{The diagrams for invisible Higgs decay to two dark matter
  particles and direct detection of dark matter through Higgs mediated
  nuclear recoil. }
\label{invis_dec}
\end{figure}

\begin{figure}[!h]
\centering
\includegraphics[width=.7\textwidth]{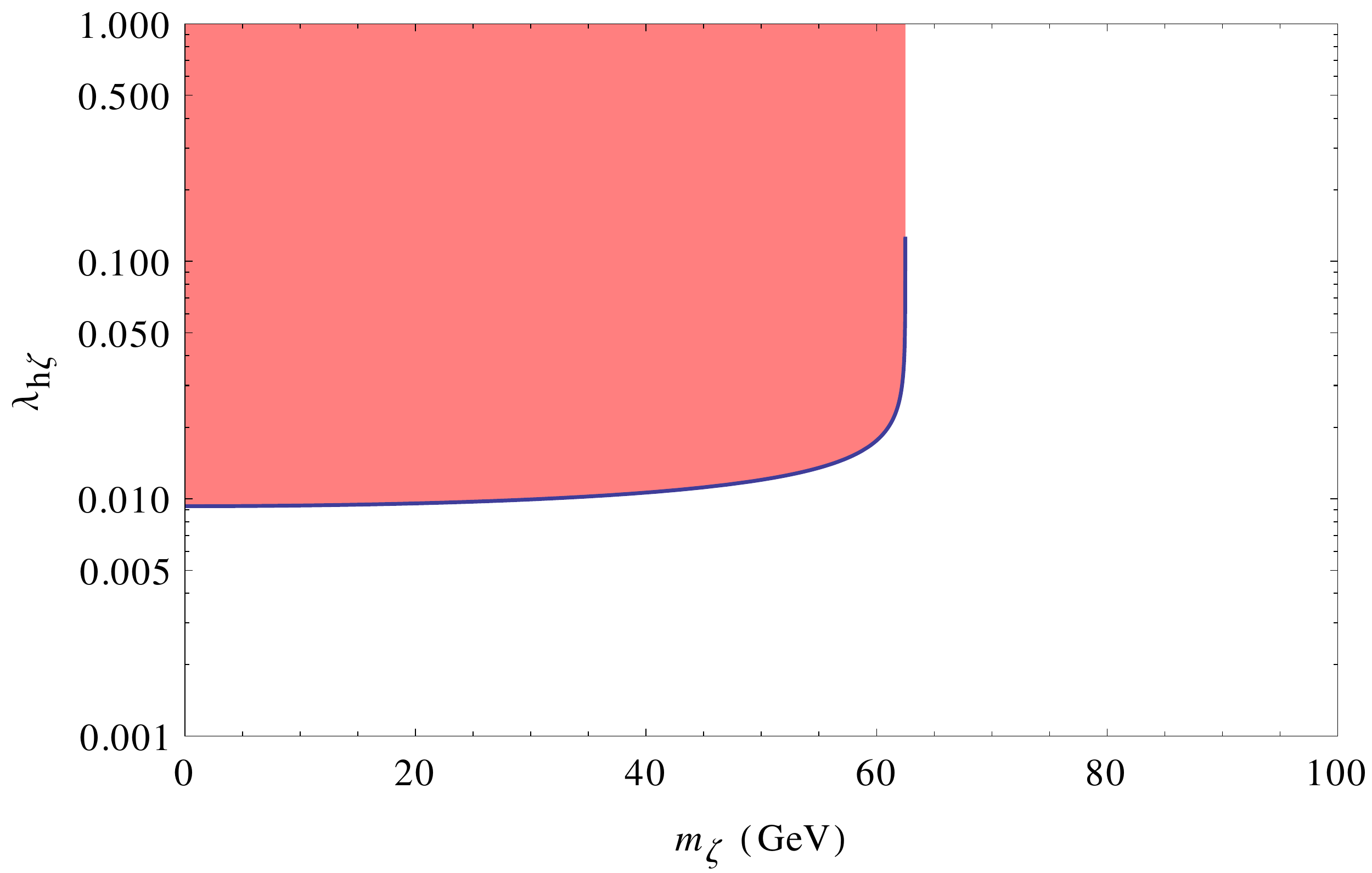}
\caption{The constraint on Higgs dark matter coupling from invisible
  Higgs decay. The shaded region is ruled out from limits on Higgs
  invisible decay width \cite{Khachatryan:2014jba}. }
\label{hdeccon}
\end{figure}

Although $\zeta$ is stable, with no direct tree level coupling to
fermions, owing to the symmetry of the model, it couples to right
handed neutrinos through exchange of $\eta$ as shown in
Fig. \ref{fig4}.
Moreover, it interacts with other scalars through cubic and quartic
terms of the type $\zeta^* \zeta \rho$ and $\zeta^* \zeta
\rho^\dagger_i \rho_j$ respectively.
In particular, the dark matter interaction with the Higgs $h$~(we
denote the recently discoverd 125 GeV particle as ``Higgs'' which in
our model will be an admixture of CP--even scalars given in
\ref{masseig}) is of special phenomenological interest.  
If the dark matter mass $m_\zeta < \frac{m_h}{2}$ then the decay $h
\to \zeta \zeta^*$, shown diagrammatically in Fig. \ref{invis_dec}, is
kinematically allowed.  The decay width of $h \to \zeta \zeta^*$ is
given by
\begin{eqnarray}
 \Gamma (h \to \zeta \zeta^*) & = & \frac{\lambda^2_{h \zeta} v^2}{16 \pi m_h} \, \sqrt{1 - \frac{4 m^2_\zeta}{m^2_h}}
 \label{hdecay}
\end{eqnarray}
where $\lambda_{h \zeta}$ is the Higgs dark matter coupling
constant. At LHC, this will be invisible decay of Higgs. Assuming no
addition beyond SM decay apart from the invisible decay, the current
LHC data puts a constraint on such invisible decay widths to be no
more than $17\%$ of the total decay width
\cite{Khachatryan:2014jba}. This puts a stringent constraint on the
Higgs dark matter coupling $\lambda_{h \zeta}$ as shown in
Fig. \ref{hdeccon}.
In our model, apart from the dark matter and other \sm Higgs decay
modes, one has decays to the new scalars $\chi, \eta$. In plotting
Fig. \ref{hdeccon}, for simplicity, we have assumed that Higgs decay
to $\chi, \eta$ is kinematically forbidden i.e. $m_\chi, m_\eta >
\frac{m_h}{2}$. It should be noted that if these decay modes are also
allowed, they will also contribute to the invisible decay of Higgs as
both $\chi, \eta$ further decay only to neutrinos.

The Higgs dark matter coupling can also be used in order to detect
dark matter directly through nuclear recoil in a variety of
experiments~\cite{Boucenna:2011tj,Ma:2015mjd}. The process proceeds
through tree--level Higgs exchange, as illustrated in
Fig. \ref{invis_dec}.
Thus our $\zeta$ WIMP dark matter model realizes the so--called
``Higgs portal'' dark matter scenario.
\begin{figure}[!h]
\centering
\includegraphics[width=.7\textwidth]{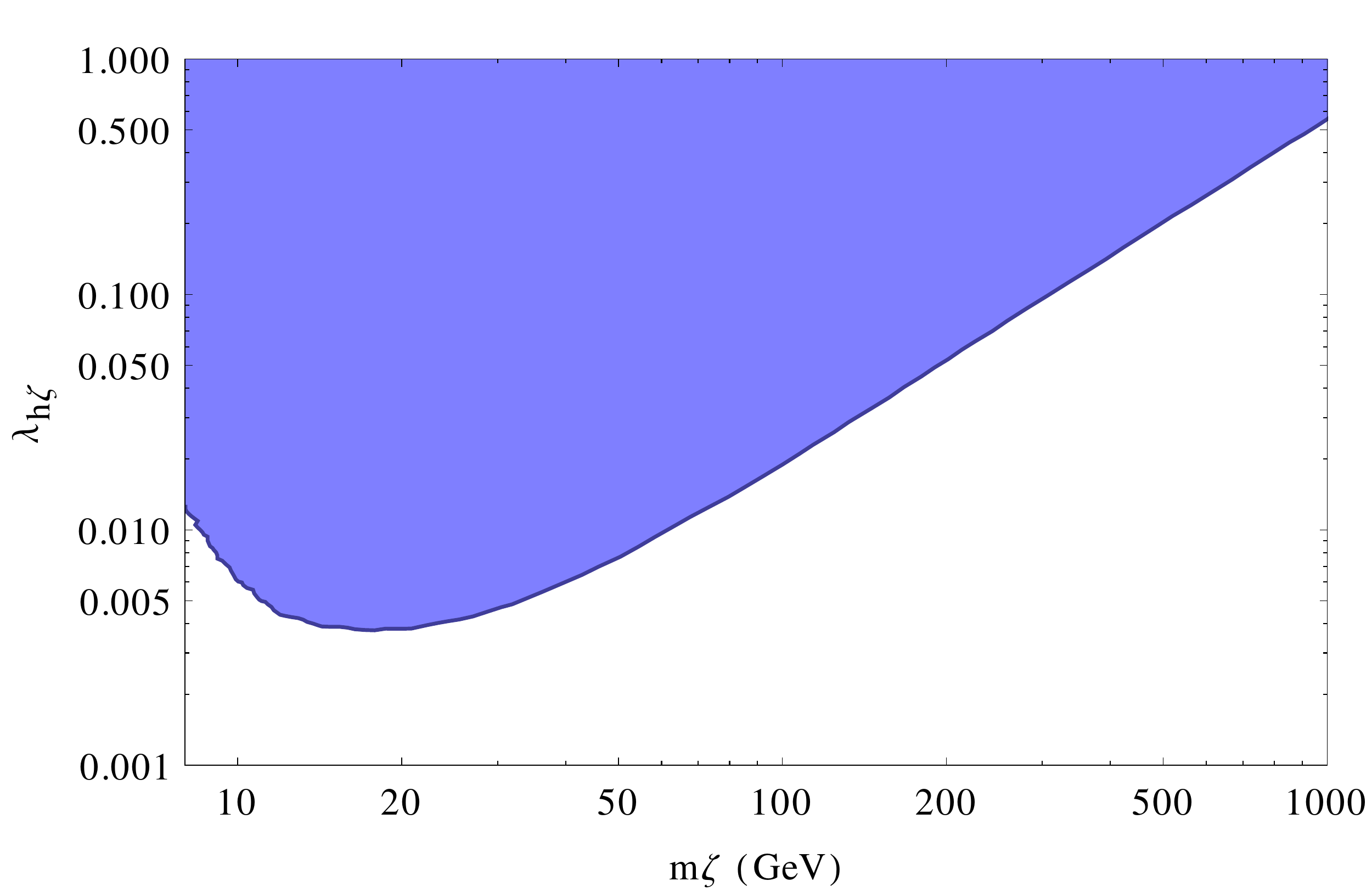}
\caption{The experimental sensitivity to our WIMP scalar dark matter
  candidate. The shaded region is ruled out by LUX data
  \cite{Akerib:2013tjd}.}
\label{cdirdet}
\end{figure}
The Higgs--mediated nucleon--dark matter cross-section in our model is
given by
\begin{eqnarray}
 \sigma & = & \frac{\lambda^2_{h \zeta} f^2_n}{\pi m^4_h} \, 
              \frac{m^4_n}{(m_n + m_\zeta)^2}
 \label{dirdm}
\end{eqnarray}
where $m_n$ is the mass of the nucleon and $f_n$ is the dimensionless
Higgs--nucleon effective coupling constant.  The null results from the
direct detection experiments such as LUX \cite{Akerib:2013tjd} lead to
rather stringent constraints on the coupling characterizing the
interaction of our WIMP scalar dark matter with the the Higgs boson.
The experimental sensitivity to our WIMP scalar dark matter candidate
is illustrated in Fig~\ref{cdirdet}. In plotting Fig~ \ref{cdirdet} we
have taken the nucleon mass $m_n = 0.93895$ GeV and the effective
Higgs nucleon coupling constant $f_n = 0.30$~\cite{Cline:2013gha}.
One sees that, in view of the region excluded by the LUX experiment,
constraints from the invisible Higgs decay $h \to \zeta \zeta$ are
relevant only in the low--mass region.

 \begin{figure}[!h]
\centering
\includegraphics[width=.8\textwidth]{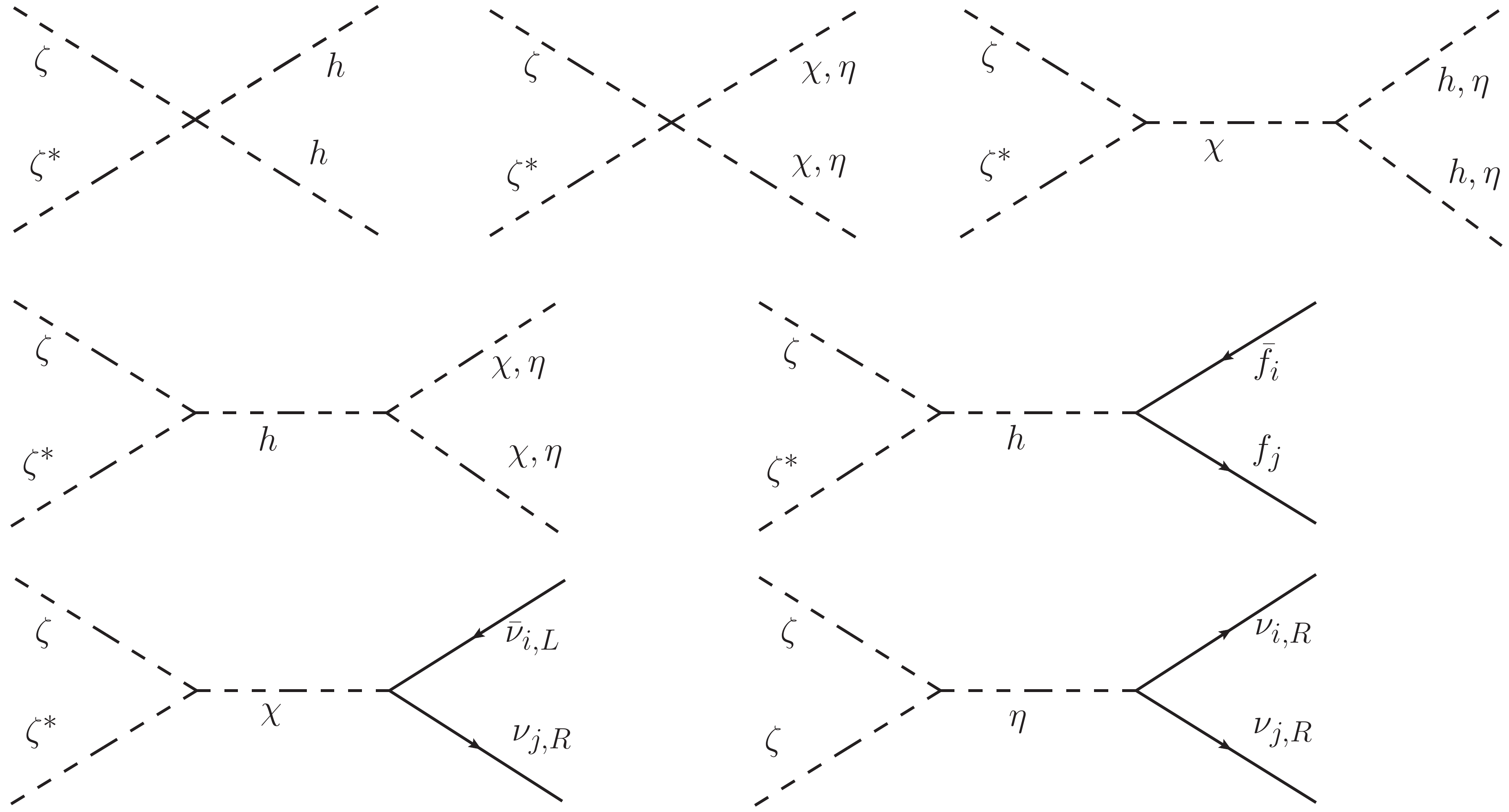}
\caption{Typical diagrams contributing to the annihilation of WIMP 
dark matter}
\label{relic}
\end{figure}

In our model we have also several terms that lead to dark matter
annihilations into two other scalars and fermions, through various
diagrams such as illustrated in Fig. \ref{relic}.
By solving numerically the relevant system of Boltzmann equations one
can show that the model parameters can be chosen to lead to the
correct relic density for dark
matter~\cite{Feng:2014vea}~\footnote{Dark matter transforming
  non-trivially under a given discrete symmetry has been previously
  studied in several works and shown indeed to provide a viable dark
  matter scenario~\cite{Boucenna:2011tj,Cline:2013gha}.  }
In addition to the direct detection signal discussed here, these terms
can also lead to indirect detection signatures associated, for
example, to gamma-ray lines from the annihilation channels $\zeta\zeta
\to \gamma\gamma$ and $\zeta\zeta \to Z\gamma$
channels~\cite{Feng:2014vea}.
A more detailed study of the discovery potential of our WIMP candidate
in direct as well as indirect detection experiments will be presented
elsewhere.

 Before ending this section let us briefly mention other features of
 our model. First, in our model a conserved $Z_4$ charge would also
 lead to the hypothetical \lnv quadruple beta decay
 process~\cite{Heeck:2013rpa}.
 Moreover, since $\eta$ is a real scalar field which couples to right
 handed neutrinos, its decay to two neutrinos or two antineutrinos may
 potentially generate a lepton asymmetry in the Universe, with
 potential application to leptogenesis with a conserved $Z_4$ lepton
 number~\cite{Heeck:2013vha}. We intend to present details of these
 and other features and implications of our model in subsequent works.
 Before closing let us also mention that our model can easily be
 generalized by including vector--like quarks, so as to accommodate
 the recent diphoton hint seen by the ATLAS and CMS collaborations. It
 would be identified with the scalar $\chi$, very much along the lines
 of Refs.~\cite{Bonilla:2016sgx,Modak:2016ung}.

\section{Discussion and Summary }
\label{sec:summary-conclusions-}

We have proposed that dark matter stability reflects the Dirac nature
of neutrinos. We illustrated how to realise this proposal within the
simplest type I seesaw mechanism for neutrino mass generation.
The scheme naturally leads to a WIMP dark matter candidate which is
made stable by the same discrete lepton number $Z_4$ symmetry, or
quarticity, which makes neutrinos to be Dirac particles. Dark matter
can be probed by searching for nuclear recoil through the Higgs portal
mechanism.
 The inclusion of the quark sector proceeds rather trivially.
Notice that, in making our point that the Diracness of neutrinos may
be responsible for dark matter stability, we have chosen a simple
``flavor-blind'' scheme based on the use of the simplest $Z_2$
symmetry to provide the seesaw mechanism.
The later can be replaced by more complex symmetries, that could also
tackle the problem of flavor and leptonic CP violation. We shall
address this issue in a follow up work.

\section*{Acknowledgements}

This research is supported by the Spanish grants
FPA2014-58183-P, Multidark CSD2009-00064, SEV-2014-0398 (MINECO) and
PROMETEOII/2014/084 (Generalitat Valenciana).

\textit{RS will like to dedicate this paper in memory of Dr. I. Sentitemsu Imsong.}


\end{document}